\documentclass[letterpaper,pre,twocolumn]{revtex4}
\usepackage{physics}
\usepackage{amsmath,amssymb}
\usepackage{epsfig}
\usepackage[geometry]{ifsym}

\newcommand{\I}{\mathrm i}

\newcommand{\A}{\mbox{A}}
\newcommand{\q}{\mbox{q}}
\newcommand{\U}{\mbox{u}}
\newcommand{\V}{\mbox{v}}
\newcommand{\suma}{\sum_{\alpha=1,2}}
\newcommand{\Om}{{\it \Omega}}
\newcommand{\ve}{\vec{\mbox{{e}}}}
\newcommand{\vect}[1]{\boldsymbol{#1}}

\begin{document}
\title{Efficient algorithms for solving the spectral scattering problems\\ for the Manakov system of nonlinear Schroedinger equations}
\date{\today}
\author{L.\,L.~Frumin$^{1,2}$}\email{lfrumin@iae.nsk.su}
\affiliation{$^{1}$Institute of Automation and Electrometry SB RAS, Novosibirsk 630090, Russian Federation,}
\affiliation{$^{2}$Novosibirsk State University, Novosibirsk 630090, Russian Federation}

\begin{abstract}
``Vectorial''  numerical algorithms are proposed for solving the inverse and direct spectral scattering problems for the nonlinear vector Schroedinger equation, taking into account wave polarization, known as the Manakov system. It is shown that a new algebraic group of 4-block matrices with off-diagonal blocks consisting of special vector-like matrices makes possible the generalization of numerical algorithms of the scalar problem to the vector case, both for the focusing and defocusing Manakov systems. As in the scalar case, the solution of the inverse scattering problem consists of inversion of matrices of the discretized system of Gelfand-Levitan-Marchenko integral equations using the Toeplitz Inner Bordering algorithm of Levinson's type. Also similar to the scalar case, the algorithm for solving the direct scattering problem obtained by inversion of steps of the algorithm for the inverse scattering problem. Testing of the vector algorithms performed by comparing the results of the calculations with the known exact analytical solution (the Manakov vector soliton) confirmed the numerical efficiency of the vector algorithms.
\medskip
\end{abstract}

\maketitle
\section{Introduction}

The nonlinear Schroedinger equation (NLSE) widely used in modern science and technology as one of the most fundamental mathematical models. NLSE belongs to the nontrivial class of integrable nonlinear partial differential equations whose solutions can be found by the Inverse Scattering Transform method (IST) \cite{NMPZ84, AS81, ZS72}.
The scalar NLSE appears abundantly in theoretical physics and nonlinear physical optics. It can also use to describe the propagation of information signals through fiber-optical communication lines \cite{MG06, KA03}.
Careful consideration of polarization phenomena in a medium with dispersion and Kerr nonlinearity is of paramount importance for the development of modern nonlinear physics and optics.

Manakov \cite{Manakov74}, when exploring self-focusing of light beams and self-induced transparency phenomena with a non-negligent contribution of polarization in nonlinear dispersive optical media, was probably the first who introduced the vector variant of the NLSE, now known as the Manakov System. This system consists of two interaction-coupled nonlinear Schroedinger equations for two optical polarizations. We will not present here the equations of the vector NLSE since the further consideration is based exclusively on the Gelfand-Levitan-Marchenko system of coupled integral equations (GLME), that applied to the solution of the spectral scattering problems in the vector case.

Manakov showed that his vector variant of the NLSE belongs to the class of integrable systems. He constructed the corresponding L-A Lax operator pair, and using the method of the IST described the general N-soliton solutions, and also found its particular solution, known as the Manakov vector soliton. Subsequently Basharov and Maimistov \cite{Basharov84} 
(see also \cite{Maimistov10}) 
discovered that the Manakov System could be used to describe many other nonlinear polarization optical effects, including the propagation of ultrashort polarized optical pulses in a resonant two-level environment.

To solve the direct and inverse scattering problems for the scalar Schroedinger equation in the frame of IST numerical algorithms of Toeplitz Inner Bordering (TIB) \cite{Belai07, Frumin15} have been developed. They based on the direct numerical solution of GLM integral equations. The TIB algorithm of the inverse scattering problem is efficient, i.e., it is fast, accurate, and stable because it is a modification of the well known Levinson algorithm \cite{Blahut85}. The TIB algorithm for solving the direct scattering problem obtained by inversion of steps of the algorithm for the inverse problem. The numerical efficiency of the algorithms caused by the Toeplitz symmetry of the discretized GLME system. TIB algorithms find applications in various optics problems, including Bragg gratings synthesis \cite{Belai07, Buryak09, Belai10}, and development of new nonlinear approaches to the transmission of information in fiber-optic lines \cite{ Aref18, Optica17, PRL17, TurOE14}.

The aim of this paper is a generalization of TIB algorithms for the solution of scattering problems for the Manakov system of vector NLSE.

In the next Section \ref{s:2}, we consider the GLM integral equations and give a short description of its application to the solution of the vector NLSE. Section \ref{s:3} describes a replacement of variables and discretization of GLME. In Section \ref{s:4}, we introduce vector-like matrices and also a new algebraic group of 4-block matrices. On the base of these constructions, in Section \ref{s:5}, we derive a vector TIB algorithm for solving the inverse scattering problem. In Section\ref{s:6}, the schematics of the vector algorithms for solving inverse and direct scattering problems are presented. Section \ref{s:7} contains some results of numerical simulation and testing of the vector TIB algorithms, and Section \ref{s:8} is a Conclusion.

\section{Vector GLME}\label{s:2}
The spectral scattering problems for the vector NLSE reduced in \cite{Basharov84} and monograph \cite{Nayanov06}) to a system of nine integral GLME, in the same way as Zakharov and Shabat did it for the scalar NLSE in the famous work \cite{ZS72}. However, this cumbersome system can split into three independent groups, each of three integral equations. It turns out that it is enough to consider the only one group of three integral equations for spectral scattering problems for the Manakov system. In the dimensionless notation close to the notation of Lam's monograph \cite{Lamb80} the system of GLME for the left-hand scattering problem consists of the next equations:
\begin{equation}\label{GLME1}\noindent
\A_0^*(x,y)+\int\limits_{-\infty}^x \suma \A_\alpha (x,z)\Omega_\alpha (y+z)dz=0,
\end{equation}
\begin{equation}\label{GLME2}
\pm \A_{\alpha}^* (x,y)+\int\limits_{-\infty}^x \A_0(x,z) \Omega_{\alpha}(y+z) dz=-\Omega_{\alpha}(x+y).
\end{equation}
Here $-x<y,~z<x$, and $\Omega_1, \Omega_2$ are GLME kernels. Here and hereinafter $ \alpha=1,2$. An asterisk means complex conjugate. The top sign of the symbol $\pm$ corresponds to the Manakov defocusing system and the lower sign to the focusing one. The difference from Lam's notation is that the indices of the functions $\A_0,\A_1,\A_2$ do not begin from unity, but from zero.

The solution of the inverse scattering problem is two components of ``potential'' vector function for two orthogonal polarizations: $\q_1 (x), \q_2 (x)$, connected with solutions of GLME equations by the synthesizing relations:
\begin{equation}\label{Synt}
\q_{\alpha}(x)=\pm 2 \A_{\alpha}^*(x,x-0).
\end{equation}

Two critical remarks should make here. First, equations \ref{GLME1} and \ref{GLME2} present actually to not two-dimensional, but three-dimensional problem, since all the considered functions $\A_0,\A_1,\A_2$ and the components of the vector kernel $\Omega_{\alpha}$ parametrically depend on an additional evolutionary variable. Since this dependence arises explicitly only when considering the evolution of the solution in time or along the optical line, it usually omitted for brevity's sake. In this paper, such an evolutionary variable corresponds to time-variable that we outline hereunder as $t$.

Secondly, for the focusing case, the components of the GLME vector kernel are the sum of the components $\mbox{V}_{\alpha} (x,t)$ of vector kernel of the continuous spectrum (the vector of pulse response function), and the sum of components $\Lambda_{\alpha}^k (x,t)$ of vector kernel for discrete eigenvalues of the Lax operator spectrum corresponding to the set of discrete eigenvalues $\{\lambda_k\}$ of the operator, where the index $k$ numbers the eigenvalues of the discrete spectrum: $\Omega_{\alpha} (x,t)= \mbox{V}_{\alpha} (x,t)+\sum_k\Lambda_{\alpha}^k (x,t).$

As in the scalar case, the solution of the Cauchy problem for vector NLSE by IST method consists of a sequence of three main steps:
\begin{enumerate}

\item	The direct spectral scattering problem. The components of known solution vector function $\q_\alpha (x,0)$, at $t=0$, are used to solve the direct scattering problem for the Manakov system, and the scattering data at $t=0$ are restored as components of the kernel vector $\Omega_\alpha (x,0)$.

\item	The spectral evolution transform of the scattering data. The components of the vector kernel $\Omega_\alpha (x,0)$, of the GLME equations at $t=0$, are transformed using the spectral evolution transform (see, for example, \cite{ZS72, NMPZ84, Lamb80}) into the components of the vector kernel GLME at time $t=T$: $\Omega_\alpha (x,T)$. Specifically, for kernel components of discrete spectrum $\Lambda_\alpha^k$, corresponding to the $k$-th eigenvalue $\lambda_k$, this transformation has the following form: $\Lambda_\alpha^k (x,T)=\Lambda_\alpha^k (x,0) \exp{(- 4 \I \lambda_k^2 T)}$, where $\I$ is imaginary unit.

\item	The inverse scattering problem. The scattering data in the form of vector kernel components $\Omega_\alpha (x,T)$, are used to solve the GLM equations and to determine the unknown components of the potential vector function $\q_\alpha (x,T)$.
\end{enumerate}

To construct a numerical algorithm, the scattering problem is considered on a finite interval $0\le x \le L$: it is assumed that the kernels $\Omega_1 (x),~\Omega_2 (x)$ vanish outside this interval. In this case, the GLM equations take the following form:
\begin{equation}\label{GLME4}\noindent
\A_0^*(x,y)+\int\limits_{-y}^x \suma \A_\alpha (x,z)\Omega_\alpha (y+z)dz=0,
\end{equation}
\begin{equation}\label{GLME5}
\pm \A_{\alpha}^* (x,y)+\int\limits_{-y}^x \A_0(x,z) \Omega_{\alpha}(y+z) dz=-\Omega_{\alpha}(x+y),
\end{equation}
where $ -x<y,~z<x\le L$, and $\alpha=1,2$.

If we put $\Omega_2 (x)=0$, then the problem becomes scalar for the potential $\q=\q_1 (x)$, and the reduced system of GLME should coincide with that for the scalar case. If we put $\Omega_1 (x)=0$, the scalar system of GLME holds similarly for the potential $\q_2 (x)$, and that is a check for the correctness of our system of GLM equations.

\section{GLME Discrete Approximation}\label{s:3}
The first step in the GLME discretization is a replacement of variables in GLME. It makes it possible to obtain integral equations with different arguments of the kernels that give matrix blocks with Toeplitz symmetry after the discretization of these equations.

Following \cite{Belai07, Frumin15} we carry out the complex conjugation of equation (\ref{GLME4}) and replace the variables: $z \rightarrow \tau-x,~y \rightarrow x-\sigma$, with $0 \le \sigma,\;\tau < 2x \le 2L $.
In equations (\ref{GLME5}), we similarly replace $y\rightarrow \tau-x,~z \rightarrow x-\sigma$.
We make also the replacement of unknown functions:
\begin{equation}\label{introduceuv}
\U(x,\sigma)=\A_0 (x,x-\sigma), \V_\alpha (x,\tau)=\pm\A_\alpha^* (x,\tau-x).
\end{equation}
Given these notations, we rewrite (\ref{GLME4}) and (\ref{GLME5}) in the form:
\begin{equation}\label{GLME7}
\U(x,\sigma)\pm \int\limits_\sigma^{2x} \suma \V_\alpha (x,\tau)\Omega_\alpha^* (\tau-\sigma)d\tau = 0.
\end{equation}
\begin{equation}\label{GLME8}
\V_\alpha (x,\tau)+\int\limits_0^\tau \U(x,\sigma) \Omega_\alpha (\tau-\sigma)d\sigma=-\Omega_\alpha (\tau).
\end{equation}
Also we respectively rewrite synthesizing relations (\ref{Synt}) as
\begin{equation}\label{Synt9}
\q_\alpha(x)=\V_\alpha (x,2x-0).
\end{equation}

We will discretize equations (\ref{GLME7}) and (\ref{GLME8}) with the 1st order of approximation accuracy. Let us introduce a discrete computational grid:
\begin{eqnarray}
&& x_m=mh/ 2;\;\; m=0,...,N;\;\; h=2L/N;\\
&& \sigma_k=k h;\; \tau_n=n h;\; k,n=0,...,m.\nonumber
\end{eqnarray}
We replace the integral $\int\limits_\sigma^{2x}\V_\alpha (x,\tau) \Omega_\alpha^* (\tau-\sigma)d\tau$ in the equation (\ref{GLME7}) by the left Riemann sum:
\begin{equation}\nonumber
\sum_{n=k}^{m-1} \V_\alpha (x_m,\tau_n ) h \Omega_\alpha^* (\tau_n-\sigma_k )
= \sum_{n=k}^{m-1} v^{(m)}_{\alpha; n} h \Om_{\alpha; n-k}^* .
\end{equation}
Here italic letters denotes the grid vectors: $v^{(m)}_{\alpha;n}=\V_\alpha (x_m,\tau_n )$ and $ \Om_{\alpha; n-k}^*= \Omega_\alpha^* (\tau_n-\sigma_k )$. Since $n \ge k$, this sum is the multiplication of the upper triangular Toeplitz matrix $Q_\alpha$ (hereinafter, we will denote ordinary matrices by capital italics) with elements $Q_{\alpha; k,n} = h\Om_{\alpha;n-k}^*$, with size $m \times m$, by the vector $v^{(m)}_{\alpha;n}$, with size $m$. The superscript $(m)$ ascribed to vector components here indicates that it corresponds to $m$th step of the algorithm. For brevity sake we do not use this superscript for matrices.
The discrete analog of equation (\ref{GLME7}) now can be represented as:
\begin{equation}\label{GLME11}
u^{(m)}_{k}\pm \sum_{n=k}^{m-1} \suma Q_{\alpha; k,n} v^{(m)}_{\alpha;n}\rm=0,
\end{equation}
where $k=0,...,m-1$. Case $m=0$ corresponds to the initial condition: $u^{(0)}_0=0$

The integrals $\int\limits_0^\tau \U(x,\sigma) \Omega_\alpha (\tau-\sigma)d\sigma$ in equations (\ref{GLME8}) can be represented using the left Riemann sum as products of low triangular Toeplitz matrices $R_\alpha$ with size $m \times m$ and with elements $R_{\alpha; n,k} = h \Om_{\alpha;n-k}$, where $n \ge k$ , on the vector $u^{(m)}_{k}=\U(x_m,\sigma_k )$.
Note here that matrices $Q_\alpha$ are Hermitian conjugations of matrices $R_\alpha$:
$Q_\alpha=R^\dagger_\alpha$.
We write the discrete analog of equations (\ref{GLME8}) in the next form:
\begin{equation}\label{GLME12}
v^{(m)}_{\alpha;n}+\sum_{k=0}^{n-1}R_{\alpha; n,k} u^{(m)}_{k}= r_{\alpha,n},
\end{equation}
where right part is $r_{\alpha,n}=-\Om_{\alpha,n}$, and $ n=0,...,m-1$.
Case $m=0$ in the form $v^{(0)}_{\alpha;0}=-\Om_{\alpha,0}$ again corresponds to the initial condition for the potential vector:
$q_{\alpha;0}=2v^{(0)}_{\alpha;0}=-2\Om_{\alpha,0}$

With $m$ changing from $1$ to $N$ we get $N$ systems of linear equations (\ref{GLME11})--(\ref{GLME12}) with size of $3m \times 3m$. These systems are ``nested'' one into another that resembles a bordering numerical algorithm. For the numerical solution of the inverse scattering problem, it is necessary to solve all the obtained nested systems and determine the components of the potential vector:
\begin{equation}\label{eq15}
q_{\alpha;m}=2v^{(m)}_{\alpha;m},~~m=1,...,N.
\end{equation}

The direct numerical solution of the nested systems of equations (\ref{GLME11})--(\ref{GLME12}) by the Gauss elimination method requires $O\left( (3N)^4 \right) $ floating-point operations, and for actual problem sizes when $N$ can reach several thousand, it is possible only with supercomputers or computer clusters. The best variant of the algorithm for solving such a series of nested linear systems seems to be a Levinson-type bordering algorithm \cite{Blahut85}, which in the process of this bordering addresses all the systems and required only $O(N^2)$ floating-point operations.

Recall that in the case of a scalar NLSE, the system of GLME consists of two coupled integral equations. The discrete form of GLME for the scalar case with first-order approximation accuracy is derived from equations (\ref{GLME11}) and (\ref{GLME12}) if we put $R_2=0$ and $v_2 =0$, and also omit the lower indices of the matrix $R_1$ and grid vector $v^{(m)}_1$. The discrete form of GLME for the scalar case has the matrix form of nested systems of linear equations:
\begin{equation}\label{BlockSystem}
\left(
\begin{array}{cc}
E &\pm R^\dagger \\
R & E \end{array} \right)
\left( \begin{array}{c}
u^{(m)}\\
v^{(m)}\end{array} \right)= \left( \begin{array}{c}
\textit{\small{0}}\\
r^{(m)}\end{array} \right),
\end{equation}
where, $m=1,..,N$, and the unknown column vector of size $2m$ is composed of two concatenated column vectors $u^{(m)}, v^{(m)}$, each of size $m$.
Block $E$ is the identity matrix, $R$ the lower triangular matrix and Hermitian conjugate $R^\dagger$ is upper triangular Toeplitz matrix, all these of a size of $m \times m$. The $\textit{\small{0}}$ in the right-hand side of Eq.~(\ref{BlockSystem}) denotes a zero column vector with size $m$, and column vector $r^{(m)}$ is given by the vector of discrete samples of the GLME kernel:
\begin{equation}\label{rvector}
r^{(m)}=-( \Om_0,...,\Om_{m-1})^T.
\end{equation}

The matrix of the system (\ref{BlockSystem}) in the scalar case has the form of a four-block matrix with Toeplitz symmetry, enabling us to apply the Levinson-type TIB algorithm for its solution.

In the case of the vector NLSE, the discretized GLM equations (\ref{GLME11}) and (\ref{GLME12}) consists of three equations. It leads to a $3m \times 3m$ block matrix consisting of nine Toeplitz blocks:
\begin{equation}\nonumber
\left(
\begin{array}{ccc}
E &\pm R_1^\dagger &\pm R_2^\dagger\\
R_1 & E&\textit{\large{0}} \\
R_2 & \textit{\large{0}} & E\end{array} \right) ,
\end{equation}
where $\textit{\large{0}}$ is zero matrix block with size $m \times m$. Despite that all these blocks are Toeplitz, the complete matrix of the system, unlike the scalar case, does not have Toeplitz symmetry. 

Note that if we rewrite vector GLME in the vector form, we get only two coupled integral equations, one scalar, and the other one vector-like. The main idea of this paper is to use the vector notation for reducing the vector case to the scalar one by presenting the discretized GLME system in the form of not nine, but four blocks, as in the scalar case, but some of them are vector-like blocks. Hereunder we introduce new mathematical constructions of the vector-like matrices and also 4-block matrices with off-diagonal vector-like matrices' that has the algebraic group properties and can be used to construct an efficient "vector" algorithm for solving the inverse scattering problem for the vector NLSE, similar to the TIB algorithm for the scalar case.

\section{2-matrices and 4-block matrices}\label{s:4}
Consider a 2-dimensional linear vector space over the field of complex numbers ${\rm C}$, with a scalar product and two orthogonal unit vectors $\ve_1$ and $\ve_2$. Elements of this vector space will be called hereinafter c-vectors. We will denote c-vectors using arrow for it and Roman letters for its components: $\vec{{\rm c}}={\rm c}_1 \ve_1+{\rm c}_2 \ve_2$.

Let us introduce the space of special vector-like matrices consisting of pairs of square $m \times m$ matrices $B_1, B_2$, that for brevity sake will be called 2-matrices: $\vect{B}=B_1 \ve_1+B_2 \ve_2$.
We will denote the 2-matrices in bold italics. We call the matrices $B_1$ and $B_2$ the projections of the 2-matrix $\vect {B}$ onto the unit vectors $\ve_1$ and $\ve_2$. Note that the $i,j$-th element of the 2-matrix $\vect{B}$ has a form $B_{1;i,j}\ve_1+B_{2;i,j} \ve_2$. The 2-matrix $\vect {B}$ is scalar-wise multiplied by the two-dimensional c-vector $\vec{b}={\rm b}_1 \ve_1+{\rm b}_2 \ve_2$ (here ${\rm b_1, b_2}$ are complex numbers), the result is the ordinary matrix ${\rm b}_1 B_1+{\rm b}_2 B_2$.

Let us define the left multiplication operation of matrix $A$ by 2-matrix $\vect{B}$: $ A\vect{B}=A B_1 \ve_1+A B_2 \ve_2$, as well as the right multiplication of 2-matrix $\vect{B}$ on the matrix $A$: $\vect{ B}A=B_1 A \ve_1+B_2 A \ve_2$. The result in both cases will be a 2-matrix. Since matrix multiplication is not a commutative operation, the results may be different.

Now we consider the scalar product of two 2-matrices $\vect{A}=A_1 \ve_1+A_2 \ve_2$ and $\vect{B}=B_1 \ve_1+B_2 \ve_2$. The result is the usual (ordinary) matrix: $C=\vect{A} \cdot \vect{B} = A_1 B_1+A_2 B_2$ (the dot $\cdot$ hereinafter denotes a scalar product). This scalar product can also be non-commutative. However, scalar multiplication is associative: $\vect{A} C \cdot \vect{B}=\vect{A}\cdot C \vect{B}$, where the $C$ matrix is ordinary.

We also introduce 2-vectors, the two projections of which are normal $m$ sized vectors, rows, or columns, which we denote in bold italics. The projections, grid vectors, we have already indicated above with simple italics. The fact that the indices $ i, k, j$, and the size $m$ also indicated in plain italics should not lead to confusion. For example, we consider a column 2-vector $\vect {b} = b_1 \ve_1 + b_2 \ve_2$ with projections $b_1, b_2$.

Note that the $k$th component of 2-vector is a c-vector and we will denote it as: $\vec{{\rm b}}_k=b_{1;k} \ve_1+b_{2;k} \ve_2$

The ordinary matrix $A$ can be multiplied on the left by the column vector $\vect{b}$, and we get the matrix form for two systems of $m$ linear equations: $A\vect{b}=\vect{c}$: $A b_1=c_1,~~ A b_2=c_2$, where 2-vectpr $\vect{c}=c_1 \ve_1+c_2 \ve_2$.

We can also multiply scalar-wise the 2-matrix $\vect{B}$ on the column 2-vector $\vect{b}$: $\vect{B} \cdot \vect{b}=(B_1 \ve_1+B_2 \ve_2) \cdot ( b_1 \ve_1+b_2 \ve_2) = B_1 b_1+B_2 b_2$. The corresponding linear equation $\vect{B}\cdot \vect{b}=d$, where $d$ is an (ordinary) vector of size $m$, can be interpreted, for example, as a matrix notation of the sum of two systems of $m$ linear equations. Left multiplication of the 2-matrix $\vect{B}$ by the (ordinary) column vector $c$ gives us, as the result, 2-vector $\vect{f}$: $\vect{B}c=\vect{f}$, ( $\vect{f}=f_1 \ve_1+f_2 \ve_2$), that should be interpreted (without discussion of their compatibility) as a compact vector representation of two linear systems:$ B_1 c=f_1 , \;\; B_2 c=f_2$.
Similarly, one can define the operations of the right multiplication of 2-row vectors on 2-matrices.

Some known properties of ordinary matrices can be extended to the 2-matrices. In particular, the 2-matrix $\vect{B}$ can be complexly conjugated by conjugating its projections: $\vect{B}^*=B_1^* \ve_1+B_2^*\ve_2$, transposed: $\vect{B}^T=B_1^T \ve_1+B_2^T \ve_2$, Hermitian conjugated: $\vect{B}^\dagger=B_1^\dagger \ve_1+B_2^\dagger \ve_2$. We define an (anti-) Hermitian 2-matrix if both its projections are (anti-) Hermitian: $B_1^\dagger = \mp B_1, B_2^\dagger = \mp B_2$. The same is applicable and for 2-vectors. We also define a Toeplitz 2-matrix if both of its projections are Toeplitz. Finally, a 2-matrix can be persymmetric if the equality $J \vect{B} J=\vect{B}^T$, holds, where $J$ is an $m \times m$ exchange matrix. We will also consider the zero 2-matrix $\vect{O}$, both projections of which are zero matrices. Note that the determinant and inversion of 2-matrix are not defined.

It is well known that ordinary non-singular matrices form a group with respect to the operation of matrix multiplication. This group has a unit element, it is the unit matrix, and an inverse element, it is the inverse matrix. The mentioned group properties of ordinary non-singular matrices allowed successfully use them for solving linear systems of equations. Vector matrices and their generalization, multidimensional matrices, have not found the same acceptance in applied mathematics and mathematical physics as ordinary matrices do, possibly because they do not have the necessary group properties. In particular, 2-matrices do not form a group with respect to the operation of generalized (scalar matrix) multiplication, since such multiplication results not in a 2-matrix, but in an ordinary matrix. Since the set of 2-matrices do not form a group, it has not a unit 2-matrix and, accordingly, has not inverse 2-matrices.
It turns out, however, that the group with respect to the generalized operation of multiplication, including ordinary matrix multiplication, multiplication of ordinary matrices by 2-matrices, and scalar multiplication of 2-matrices, forms a more complex construction of 4-block matrix whose diagonal blocks are formed by ordinary matrices, for example, $A$ and $B$, and off-diagonal blocks are 2-matrices, for example, $\vect{B}$ and $\vect{C}$ with projections sizes of $m \times m$:
\begin{equation}\label{Mmatrix}
\mathbf{M} = \left( \begin{array}{cc}
A & \vect{B} \\
\vect{C} & D \end{array} \right),
\end{equation}
Hereinafter, we denote such 4-block matrices by capital Roman bold letters and will also call simply block matrices. For such block matrices, there is a unit element: $\mathbf{E} = \left( \begin{array}{cc}
E & \textbf{\textit{\large{0}}} \\
\textbf{\textit{\large{0}}} & E \end{array} \right)$,
where $E$ is the unit matrix of size $m\times m$, and $\textbf{\textit{\large{0}}}$ is the 2-matrix with zero projections of $m \times m$ size. For non-singular (it will be clear later what it means) 4-block matrices there exists an inverse matrix having the same 4-block form.
We write the generalized Frobenius formula \cite{FRG00,Bernsrein05} for the inversion of our $\mathbf{M}$ block matrix (\ref{Mmatrix}):
\begin{equation}\label{Frobenius}
\mathbf{M}^{-1}=\left( \begin{array}{cc}
A^{-1}+A^{-1}\vect{B}\cdot H\vect{C}A^{-1} &- A^{-1}\vect{B}H \\
-H\vect{C}A^{-1} & H \end{array} \right),
\end{equation}
where $H = (D - \vect{C}A^{-1}\cdot \vect{B})^{-1}$.
It can also see from (\ref{Frobenius}) that the inverse matrix has ordinary matrices on its diagonal, and the off-diagonal blocks are 2-matrices. We emphasize that the Frobenius formula does not require the inversion of the diagonal 2-matrices $\vect{B}$ and $\vect{C}$, for which the inversion not defined. It follows from the Frobenius formula that an inverse matrix exists if there exist matrices $A^{-1}$ and $H$. Besides, the scalar product of 2-matrices is associative to ensure the equality of the left and right inverse block matrices.

The fact that non-singular block matrices form a group with respect to the generalized multiplication operation allows us to use them to solve systems of linear equations, and also expand the range of applicability of some numerical algorithms and approaches developed for ordinary non-singular matrices.

To compose a linear system of equations with a block matrix we consider a block (column) vector $\mathbf{p}$, denoted by a Roman bold letter, that is an analog of columns from the left half of the block matrix. Its top part is an (ordinary) column vector $c$, with $m$ size, and the bottom part is a 2-vector column $\vect{b}$: $\mathbf{p}=\left( \begin{array}{c} c\\ \vect{b} \end{array} \right)$. It is easy to verify, following the rules described above for multiplying ordinary and 2-matrices by an ordinary and 2-vector, that multiplying a unit block matrix $\mathbf{E}$ by a block vector $\mathbf{p}$ leaves the latter unchanged: $\mathbf{E p}=\mathbf{p}$. There is another version of the block vector, it is a ``flip'' block vector, that corresponds to the columns of the right half of the block matrix, for example, $\mathbf{d}=\left( \begin{array}{c} \vect {b} \\ c \end{array} \right)$.
It is also easy to verify that $\mathbf{E d}=\mathbf{d}$. Similarly one can define block row vectors.

\section{Vector TIB algorithm for the inverse scattering problem}\label{s:5}
We turn to the system of equations (\ref{GLME11}) (\ref{GLME12}) using the block vector notations described above. First we consider 2-matrices $\vect{R}=R_1 \ve_1+R_2 \ve_2$, where $R_1, R_2$ are Toeplitz $m \times m$ matrices, $m=1,...,N$.
Also we define a 2-vector column of the solution of the system of equations (\ref{GLME11}) and (\ref{GLME12}) $\vect{v^{(m)}}=v^{(m)}_1 \ve_1+v^{(m)}_2 \ve_2$, and a 2-vector column of the right-hand side $\vect{r}^{(m)}=r^{(m)}_1 \ve_1+r^{(m)}_2 \ve_2$, where the column vectors $\vect{r}^{(m)}$ are given by the discretized kernels of the GLM equations, as in the scalar case in Eq. (\ref{rvector}). It is required to find the 2-vector of the potential $\vect{q}=q_1 \ve_1+q_2 \ve_2$. We denote its $n$th component as c-vector $\vec{q}_n$. In these notations, equations (\ref{GLME11}) and (\ref{GLME12}) like in scalar case (\ref{BlockSystem}) can be represented in the form of a 4-block matrix:
\begin{equation}\label{VBlockSystem}
\left(
\begin{array}{cc}
E &\pm \vect{R}^\dagger \\
\vect{R} & E \end{array} \right)
\left( \begin{array}{c}
u^{(m)}\\
\vect{v}^{(m)}\end{array}
\right)= \left( \begin{array}{c}
{\it 0}\\
\vect{r}^{(m)}\end{array} \right)
\end{equation}
From a comparison of (\ref{VBlockSystem}) and (\ref{BlockSystem}) it follows that the vector case differs from the scalar one only by the corresponding vector notation. The matrix of system (\ref{VBlockSystem}) is precisely a 4-block matrix of size $2m \times 2m$, the diagonal blocks of which are formed by ordinary unit matrices $E$, and the off-diagonal blocks formed by 2-matrices $\vect{R}$ and $\pm \vect{R}^\dagger$. Since the latter are Toeplitz 2-matrices, the entire block matrix of the system
$\mathbf{G} = \left( \begin{array}{cc}
E &\pm \vect{R}^\dagger \\
\vect{R} & E \end{array} \right)$,
is also Toeplitz. Off-diagonal blocks of this matrix have either Hermitian or anti-Hermitian symmetry, depending on the sign $\pm$, which makes it possible to develop the vector versions of the TIB algorithms. The Levinson algorithm for Toeplitz matrices is not directly applicable to this problem. As in the scalar case, if we increase index $m$ by one an ``inner bordering'' occurs for 2-matrices $\vect{R}$, $\pm\vect{R}^\dagger$ and also for identity matrices $E$, each of them increases by one column and one row. The block matrix $\mathbf{G}$, in this case, increased by two rows and two columns, while in the Levinson algorithm at each step, the size of the matrix incremented by 1. In this case, the analog of the TIB algorithm for the inverse scattering problem, described in \cite{Belai07, Frumin15} becomes applicable.

Suppose that at the $m$th step of the algorithm we know the solution in the form of a column block vector $ \left( \begin{array}{c}
u^{(m)}\\
\vect{v}^{(m)}\end{array}\right)$. It is required at the next step of the algorithm to find the solution $\vect{v}^{(m+1)}$ corresponding to the embedded system of equations of size $2m+2$.

We will obtain the solution of the system of linear equations if we find the inverse block matrix $\mathbf{G}^{-1}$. Using the generalized Frobenius formula (\ref{Frobenius}), we write formally this inverse matrix in the form:
\begin{equation}\label{GFrobenius}
\mathbf{G}^{-1}=\left( \begin{array}{cc}
E \pm \vect{R}^\dagger H \cdot\vect{R} & \mp \vect{R}^\dagger H \\
-H \vect{R} & H \end{array} \right),
\end{equation}
where $H=(E \mp \vect{R}\cdot \vect{R}^\dagger )^{-1}$. Note that matrix $E \mp \vect{R}\cdot \vect{R}^\dagger$ is Hermitian; therefore, matrix $H$ is also Hermitian. In addition, the top diagonal block $E \pm \vect{R}^\dagger H \cdot\vect{R}$ of $\mathbf{G}^{-1}$ is also Hermitian, i.e. both diagonal blocks of the inverse matrix are Hermitian. The off-diagonal blocks of the inverse matrix are either Hermitian (for the upper sign, i.e., for the defocusing NLSE), or anti-Hermitian (for the focusing NLSE). The symmetry properties of the blocks of the inverse matrix are important for constructing a vector algorithm similar to the scalar TIB. For this algorithm, relations between the left $\mathbf{f}^{(m)}_1$ and right $\mathbf{f}^{(m)}_m$ block column of the inverse matrix, and also its top $\mathbf{g}^{(m)}_1$ and bottom $\mathbf{g}^{(m)}_m$ block row are of particular importance.

The symmetry of blocks of the inverse matrix $\mathbf{G}^{-1}$ allows us to establish relationships between columns and rows framing the matrix. Note that to solve the inverse scattering problem it is required to find only one bottom block row $\mathbf{g}^{(m)}_m$ of the inverse 4-block matrix. The knowledge of this row is sufficient to determine the $m$th component of the 2-vector of potential $\vec{q}_m$ since it is determined by only one last element $\vec{v}^{(m)}_m$ (it is c-vector) of the solution 2-vector $\vec{v}^{(m)}$: $\vec{q}_m=2\vec{v}^{(m)}_m = 2\mathbf{g}^{(m)}_m \left( \begin{array}{c}
{\it 0}\\
\vect{r}^{(m)}\end{array} \right) $.

For what follows, it is convenient to introduce for $m$th step of the algorithm the column vector $y^{(m)}$ and the column 2-vector $\vect{z}^{(m)}=z^{(m)}_1 \ve_1+z^{(m)}_2 \ve_2$.
Let the first (left) block column $\mathbf{f}_1$ of the inverse matrix consist of these vectors:
$\mathbf{f}^{(m)}_1= \left( \begin{array}{c}
y^{(m)}\\
\vect{z}^{(m)}\end{array} \right)$.
The Toeplitz symmetry of the original $\mathbf{G}$ matrix ensures the persymmetry of the inverse matrix. Therefore, the bottom block row $\mathbf{g}^{(m)}_m$ of the inverse matrix $\mathbf{G}^{-1}$ is a symmetric reflection of the left block column $\mathbf{f}^{(m)}_1$ with respect to the northeast-to-southwest diagonal: $\mathbf{g}^{(m)}_m = \left( \begin{array}{c}\widetilde{\vect{z}^{(m)}} \\
\widetilde{y^{(m)}}\end{array} \right)^T$, where the tilde means the inverse of the numbering of the elements of a row or column. Due to the Hermitian symmetry of the diagonal blocks of the inverse matrix, the left part of the top half of the block row $\mathbf{g}^{(m)}_1$ of the inverse matrix, that is part of its top diagonal block, is Hermitian conjugate to the first half of the left column $y^{(m)}$ and has the form $y^{(m)\dagger}$. For off-diagonal blocks of the inverse matrix in the Hermitian/anti-Hermitian cases, the 2-vector row of the top block row $\mathbf{g}^{(m)}_1$ is Hermitian or anti-Hermitian conjugate to the left column vector $\vect{z}^{(m)}$. Thus, the top block row of the inverse matrix can be represented as $\mathbf{g}^{(m)}_1=\left( \begin{array}{c} y^{(m)}\\
\pm\vect{z}^{(m)}\end{array} \right)^\dagger$.
Taking into account the persymmetry of the inverse matrix, we see that the right (last) block column $\mathbf{f}^{m)}_m$ of the inverse matrix is persymmetric to the top (first) row $\mathbf{g}^{(m)}_1$, that is $\mathbf{f}^{(m)}_m=\left( \begin{array}{c}
\pm\widetilde{\vect{z}^{(m)}}\\
\widetilde{y^{(m)}}\end{array} \right)^\dagger$.

Comparing the obtained rows and columns of inverse matrices for the scalar and vector cases of the GLME, we come to a vector generalization of the TIB inverse scattering algorithm, which differs from the scalar one only in that some of the vector arrays (for example, $y^{(m)}$) in the algorithm remain the same, but another one($\vect{z}^{(m)}$) becomes 2-vector, i.e. as if they became doubled. The auxiliary vectors $y^{(m)}$ and $\vect{z}^{(m)}$ at $m$th step of the algorithm are calculated on the basis of the equations:
\begin{equation}\label{ystep}
y^{(m+1)}=c^{(m)}\left( \begin{array}{c}
y^{(m)}\\
0 \end{array} \right)+\vec{d}^{(m)} \cdot \left( \begin{array}{c}
\vec{0}\\
\pm\widetilde{\vect{z}^{(m)*}} \end{array} \right)
\end{equation}
\begin{equation}\label{zstep}
\vect{z}^{(m+1)}=c^{(m)}\left( \begin{array}{c}
\vect{z}^{(m)}\\
\vec{0} \end{array} \right)+\vec{d}^{(m)} \left( \begin{array}{c}
0\\
\pm\widetilde{y^{(m)*}} \end{array} \right)
\end{equation}
Here $\vec{0}$ is a c-vector with two zero components, $\mbox{c}^{(m)}$ is a complex scalar, and $\vec{ d}^{(m)}=\mbox{d}_1^{(m)}\ve_1+\mbox{d}_2^{(m)} \ve_2$ is a c-vector. Equation (\ref{ystep}) is scalar, and (\ref{zstep}) is a vector equation, i.e. these are two equations for two components of the 2-vector $\vec{z}^{(m+1)}$.

For the $\mathbf{f}^{(m)}_1$ and right $\mathbf{f}^{(m)}_m$ block columns of the inverse block matrix, we can write:
\begin{equation}\label{Gf}
\mathbf{G f}^{(m)}_1=(1 0...0\; \vect{\it 0})^T, \;\; \mathbf{G f}^{(m)}_m=(\vect{\it 0}\; 0...0 1)^T.
\end{equation}
Here $ \vect{\it 0}$ is a zero 2-vector. Comparing equations (\ref{Gf}) for step $m$ and for step $m+1$, we arrive at the following equations for the coefficients $\mbox{c}^{(m)}, \vec{d}^{(m)}$:
\begin{equation}\label{eqforcd}
\mbox{c}^{(m)}\pm \vec{\beta}^{(m)*} \cdot \vec{d}^{(m)} = 1,\;\;\;\mbox{c}^{(m)}\vec{\beta}^{(m)} \pm \vec{d}^{(m)}=\vec{0}.
\end{equation}
Here one equation is also scalar and the other is vector, i.e. compact record of 2 equations.
The solution to the system (\ref{eqforcd}) has the form:
\begin{equation}\label{cdsolution}
\mbox{c}^{(m)} = (1 \pm \mid \vec{\beta}^{(m)}\mid ^2)^{-1},\;\;
\vec{d}^{(m)}=\vec{\beta}^{(m)}\mbox{c}^{(m)}
\end{equation}

The main parameter of the TIB algorithm c-vector $\vec{\beta}^{(m)}$ has the components: $\vec{\beta}^{(m)}=\beta^{(m)}_1 \ve_1+\beta^{(m)}_2 \ve_2$, and is given by:
\begin{equation}\label{beta}
\vec{\beta}^{(m)}=\sum_{k=0}^{m-1} h\vec{\Om}_{m-k} y^{(m)}_{k}
\end{equation}
This equation is also a compact vector representation of a pair of equations.
As a result, for the $m$th component of the 2-vector of the potential, we obtain:
\begin{equation}\label{solution}
\vec{q}_m=2\vec{v}^{(m)}_m =-2\vec{\beta}^{(m)}/h.
\end{equation}

\section{Schematic of the vector TIB algorithms }\label{s:6}
\subsection{Vector TIB algorithm for the inverse scattering problem}
First-order algorithm for the inverse scattering problem includes the following steps:
\begin{enumerate}
\item Put $m = 1$ and calculate initial value for 0th component of the solution vector (it is a c-vector) $\vec{q}_0 = -2\vec{\Om}_0$ and initial values for auxiliary vectors:
\begin{equation}\label{auxilary}
y^{(1)}_0=(1\pm h \mid \vec{\Om}_0 \mid^2 )^{-1},\;\; \vec{ z}^{(1)}_0 = -y^{(1)}_0 h \vec{\Om}_0.
\end{equation}
\item	Determine the main parameter of the algorithm c-vector $\vec{\beta}^{(m)}$ using (\ref{beta}).
\item Find $m$th component of the potential vector $\vec{q}_m$ from Eq. (\ref{solution}); this is the output at every step.
\item	Calculate coefficients $\mbox{c}^{(m)}$ and $\vec{d}^{(m)}$ from Eq. (\ref{cdsolution}).
\item	Determine auxiliary vector $y^{(m)}$ and 2-vector $\vect{z}^{(m)}$ using (\ref{ystep}) and (\ref{zstep}).
\item Increment $m$ and go to the step 2 until $m<N$.
\end{enumerate}

\subsection{Vector TIB algorithm for the direct scattering problem}
Described above algorithm for the inverse scattering problem can be inverted to solve the direct scattering problem. Resulting algorithm consists of the following steps:
\begin{enumerate}
\item	Calculate initial values the kernel vector $\vec{\Om}_0=-\vec{q}_0/2$, and initial values for the auxiliary vectors (\ref{auxilary}) and put $m=1$.
\item	Determine the main parameter of the algorithm: $\vec{\beta}^{(m)}=-h\vec{q}_m/2$.
\item	Find $m$th component of the kernel vector $\vec{\Om}_m$ (this c-vector is the output at every step):
\begin{equation}\nonumber
\vec{\Om}_m=\left(\vec{\beta}^{(m)} - h \sum_{k=1}^{m-1} \vec{\Om}_{m-k} y^{(m)}_{k}\right)/y^{(m)}_0.
\end{equation}
\item	Calculate coefficients $\mbox{c}^{(m)}$ and $\vec{d}^{(m)}$ from Eq. (\ref{cdsolution}).
\item	Determine auxiliary vector $y^{(m)}$ and 2-vector $\vect{z}^{(m)}$ using (\ref{ystep}) and (\ref{zstep}).
\item Increment $m$ and go to the step 2 until $m<N$.
\end{enumerate}

\section{Numerical simulation: algorithms verification}\label{s:7}
Numerical simulation was performed to test the presented vector TIB algorithms for the Manakov vector soliton as an example of an accurate solution. Recall that the Manakov vector soliton corresponds to one eigenvalue $\lambda = \omega+\I a$ of the discrete spectrum of the Manakov system:
\begin{equation}\label{ManakovSoliton}
\q_{\alpha}(x) =-2\,\mbox{l}_{\alpha} a\,\mbox{sech}(2 a x+\delta) \exp(-2 \I \omega x + \I \theta),
\end{equation}
where $\mbox{l}_\alpha$ are components of c-vector of polarization of the soliton, $a$ is amplitude, $\delta$ is its center displacement, $\omega$ is the frequency, $\theta$ is its phase, and as elsewhere in the text $ \alpha = 1,2.$

This soliton corresponds to the next components of vector kernel of the GLME
\begin{equation}\label{ManakovKernel}
\Omega_{\alpha}(x) = {\rm c}_\alpha \exp(-\I \lambda x) = {\rm c}_\alpha \exp(-\I \omega x+a x).
\end{equation}
Here ${\rm c}_\alpha = 2 \mbox{l}_\alpha a \exp{(\delta+\I \theta)}$ are components of the complex vector constant that determines amplitude, shift, phase and polarization of the soliton.

\begin{figure}\centering
\includegraphics[width=\columnwidth]{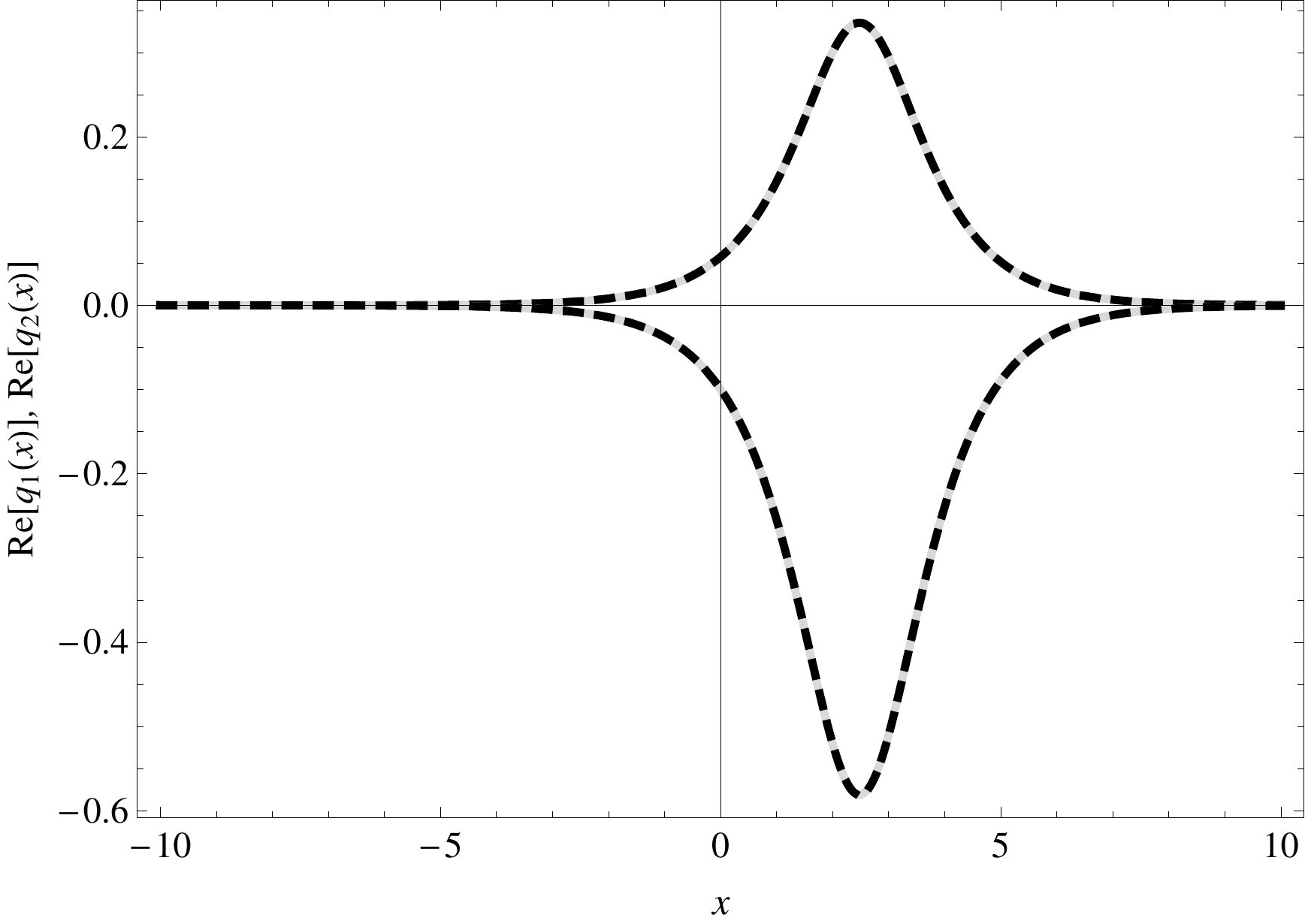}
\caption{Inverse scattering problem: comparison of the exact (gray curve) and restored (black strokes) real parts of two polarization components of the potential 2-vector $\vect{q}$ for the Manakov vector soliton. Real part of the component $q_1$ of the potential vector is placed above the abscissa axis and real part of the component $q_2$ lies below the abscissa axis.}\label{fig:1}
\end{figure}

\begin{figure}\centering
\includegraphics[width=\columnwidth]{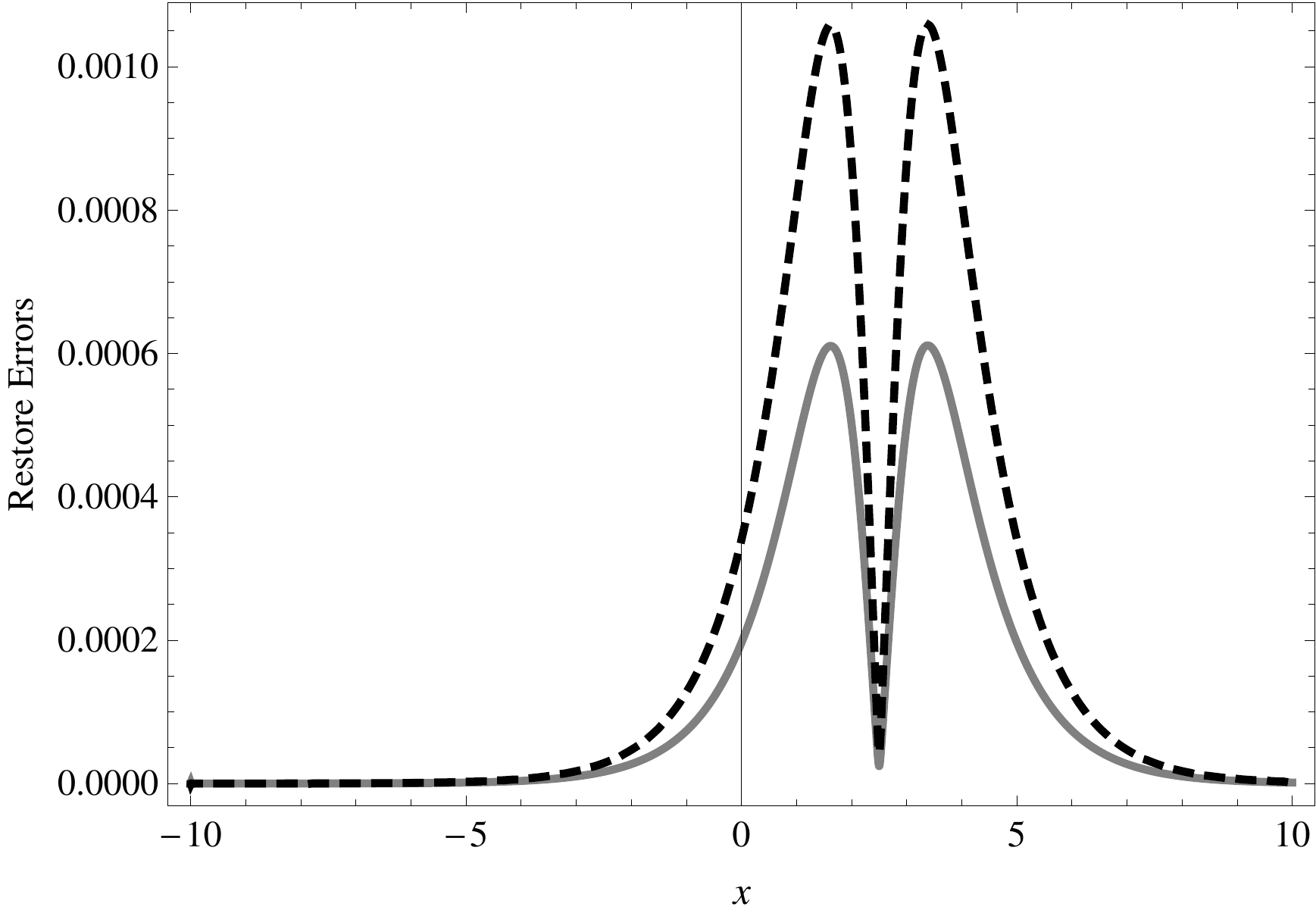}
\caption{Inverse scattering problem: distribution of absolute calculation errors for both polarization components of the potential 2-vector $\vect {q}$ of the Manakov vector soliton. Gray curve refers to the component $ q_1$ , and black strokes correspond to the component $ q_2 $.}\label{fig:2}
\end{figure}

\begin{figure}\centering
\includegraphics[width=\columnwidth]{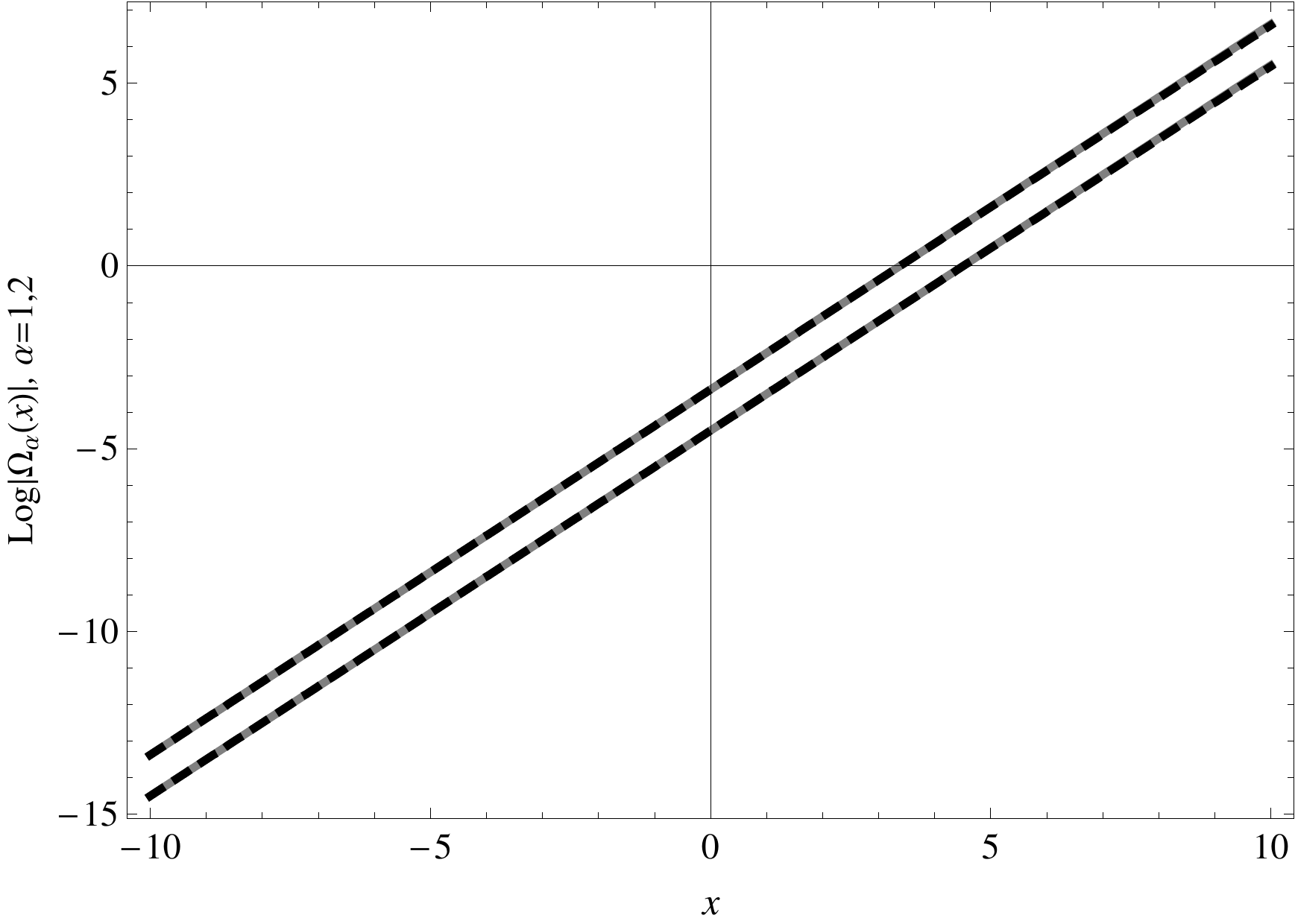}
\caption{Direct scattering problem: Comparison of the exact (gray curve) and restored (black strokes) logarithms of the absolute value of integral kernels $\Om_\alpha$ for two its polarization components.}\label{fig:3}
\end{figure}

\begin{figure}\centering
\includegraphics[width=\columnwidth]{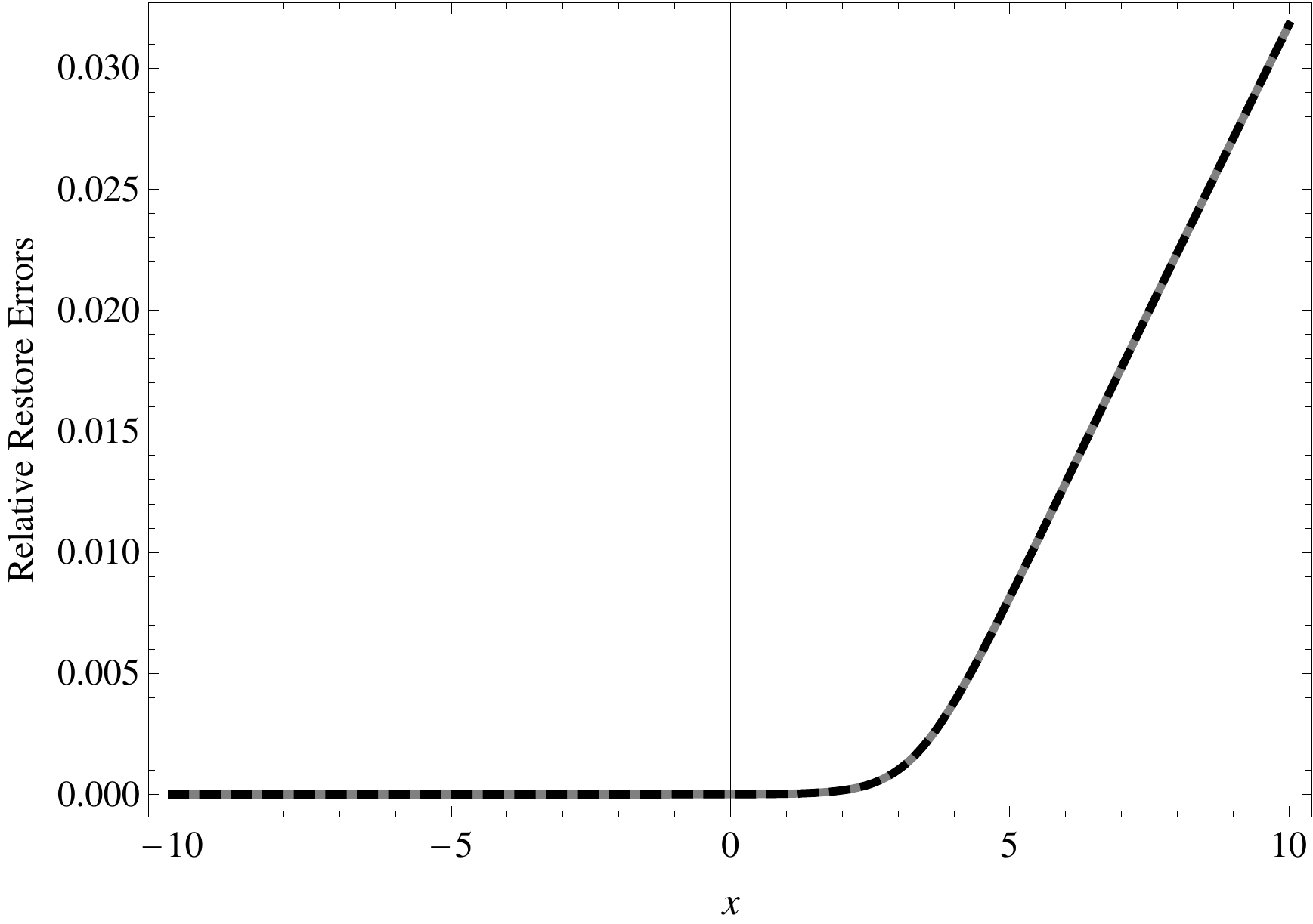}
\caption{Direct scattering problem: distribution of the relative calculation errors of integral kernels $\Om_\alpha$, $\alpha=1,2$ for both polarization components. The gray curve refers to the component $\Omega_1$, the black strokes correspond to component $\Omega_2$ of kernel vector.}\label{fig:4}
\end{figure}

Numerical modeling confirmed the efficiency of the vector TIB algorithms. Some of the calculation results, using the example of the Manakov vector soliton, are shown in Fig.~\ref{fig:1}--\ref{fig:4}. The calculations performed by the variant of the program for solving the inverse scattering problem on the interval $-L/2 \le x\le L/2$. The solution obtained on the range $[-10, 10]$ for $N=2^{13}=8192$ calculation intervals, and soliton shifted from the center of the interval to test the asymmetric solution. The polarization of the soliton chosen so that the real parts of components of the soliton potential have different signs. The exact and restored from the GLM kernel real parts of the Manakov 2-vector potential (\ref{ManakovSoliton}) are presented in the figure \ref{fig:1}. One polarization component of the Manakov soliton displayed above the abscissa axis, and below there is another one. The calculations carried out with first-order approximation accuracy.

Figure \ref{fig:2} presents the distribution of the absolute value of the solution error for the inverse scattering problem for $N=2^{13}$ calculation intervals. The maxima of the absolute error in the figure correspond to the peaks of the derivative solution. It follows that the main is the approximation error. The error fall at the end of the interval confirms that it not accumulated in the algorithm.
The integral error of inverse problem solution for $N=2^{12}$ calculation intervals was 0.0028, and for $2^{13}$ intervals, it was 0.0012. The ratio of these values is 2, which indicates the 1st order of approximation accuracy. The calculation time increased four times that correspond to $O(N^2)$ floating-point operations.

Figures \ref{fig:3} and \ref{fig:4} presents calculation result for the direct scattering algorithm for $N=2^{13}$ calculation intervals. Figures \ref{fig:3} shows a comparison of the exact and restored logarithms of the absolute value of integral kernel 2-vector $\Om_\alpha$ for both polarization components. Figure \ref{fig:4} gives the distribution of the relative calculation errors of integral kernels $\Om_\alpha$. It can see from the figure that, starting from about the middle of the Manakov soliton, the relative calculation error increases almost linearly. It indicates a moderate error accumulation by the TIB algorithm for solving the direct scattering problem.
However, when the number of calculation intervals doubled, the integral error also halves, which confirms the first order of approximation accuracy.


\section{Conclusion}\label{s:8}
Based on the discovered group properties of 4-block matrices with vector-like off-diagonal matrix blocks, a generalization of the efficient scalar TIB algorithms for solving inverse and direct spectral scattering problems for the vector nonlinear Schroedinger equation (Manakov system) is presented. Similar to the scalar algorithms, the new algorithms are based on solving the discretized system of coupled GLM integral equations for both focusing and defocusing cases. Also, as in the scalar case, the acceleration of calculations is achieved due to the Toeplitz symmetry of the matrix of the discretized system of GLM equations. The algorithms were tested on the exact NLSE analytical solution, Manakov vector soliton, and demonstrated high speed, stability, and accuracy, which is sufficient for many applications.

\section{Acknowledgements}\label{s:9}
The author is grateful to Professor D.~A. Shapiro and Professor S.~K. Turitsyn for helpful discussions, recommendations, and interest in this work.

Funding: This work was supported by the Ministry of Science and Higher Education of Russian Federation, project AAAA-A17-117062110026-3, and Section \ref{s:7} (Numerical Simulation) was supported by Russian Science Foundation (RSF) (17-72-30006).


\begin{thebibliography}{10}
\bibitem{NMPZ84}
S.~Novikov, S.~Manakov, L.~Pitaevskii, and V.~Zakharov, {\it Theory of solitons: the inverse scattering method} (Springer Science \& Business Media, 1984).
\bibitem{AS81}
M.~J.~Ablowitz and H.~Segur, {\it Solitons and the inverse scattering transform}, Vol. 4 (SIAM Studies in Applied Mathematics, Philadelphia 1981).
\bibitem{ZS72}
V. Zakharov and A. Shabat, Soviet Physics JETP {\bf 34}, 62 (1972).
\bibitem{MG06}
L.~F. Mollenauer and J.~P. Gordon, {\it Solitons in optical fibers: fundamentals and applications} (Academic Press, 2006).
\bibitem{KA03}
Y.~S. Kivshar and G. Agrawal, {\it Optical solitons: from fibers to photonic crystals} (Academic press, 2003).
\bibitem{Manakov74}
S.~V. Manakov, Soviet Physics JETP {\bf38}, 248 (1974).
\bibitem{Basharov84}
A.~M. Basharov and A.~I. Maimistov, Soviet Physics JETP {\bf 60}, 913 1984.
\bibitem{Maimistov10}
A.~I. Maimistov, Quantum Electronics {\bf 40}, 756 (2010).
\bibitem{Belai07}
O.~V. Belai, L.~L. Frumin, E.~V. Podivilov, and D.~A. Shapiro, Journal of the Optical Society of America B {\bf 24}, 1451 (2007).
\bibitem{Frumin15}
L.~L. Frumin, O.~V. Belai, E.~V. Podivilov, and D.~A. Shapiro, Journal of the Optical Society of America B {\bf 32}, 290 (2015).
\bibitem{Blahut85}
R.~E. Blahut, {\it Fast Algorithms for Digital Signal Processing} (Addison-Wesley, 1985).
\bibitem{Buryak09}
A. Buryak, J. Bland-Hawthorn, and V. Steblina, Opt. Express {\bf 17}, 1995 (2009).
\bibitem{Belai10}
O.~V. Belai, L.~L. Frumin, E.~V. Podivilov, and D.~A. Shapiro, Laser Physics {\bf 20}, 318 (2010).

\bibitem{Optica17}
S.~K. Turitsyn, J.~E. Prilepsky, S.~T. Le, S.~Wahls, L.~L. Frumin, M. Kamalian, and S.~A. Derevyanko, Optica {\bf 4}, 307 (2017).

\bibitem{Aref18}V.~Aref, S.~T. Le, and H.~Buelow IEEE Journal of lightwave technology {\bf 36}, 1289 (2018 ).

\bibitem{PRL17}
L.~L. Frumin, A. Gelash, and S.~K. Turitsyn, Physical Review Letters {\bf 118}, 223901 (2017).

\bibitem{TurOE14}
S. Le, Y. Prylepskiy, and S. Turitsyn, Optics Express {\bf 22}, 26720 (2014)


\bibitem{Nayanov06}
V.~I. Nayanov, {\it Multi-Field Solitons} (Fizmztlit, Moscow 2006, in Russian).

\bibitem{Lamb80}
G.~L. Lamb Jr., {\it Elements of soliton theory} (New York, Wiley-Interscience, 1980).

\bibitem{FRG00}
F.~ R. Gantmacher, {\it Theory of Matrices}. (AMS Chelsea Publishing: Reprinted by American Mathematical Society, 2000).

\bibitem{Bernsrein05}
D. Bernstein, {\it Matrix Mathematics} (Princeton University Press, 2005).
\end{thebibliography}
\end{document}